\documentclass[conference]{IEEEtran}
\IEEEoverridecommandlockouts

\usepackage{tikz}
\usepackage{amsmath}
\usepackage{amsmath,amssymb,amsfonts}
\usepackage{algorithmic}
\usepackage{graphicx}
\usepackage{textcomp}
\usepackage{xcolor}
\usepackage{listings}

\usepackage{cite}
\usepackage{hyperref}
\usepackage[bf, small]{caption}

\usepackage{balance}
\usepackage{booktabs} 
\usepackage{subfigure}
\usepackage{subcaption}
\usepackage{multirow}

\usepackage{xurl}
\usepackage{balance}

\usepackage{graphics,subfigure, subfloat}
\def\BibTeX{{\rm B\kern-.05em{\sc i\kern-.025em b}\kern-.08em
    T\kern-.1667em\lower.7ex\hbox{E}\kern-.125emX}}

\def\twocolfig#1#2#3#4{
\begin{figure}[t]
\begin{center}
\includegraphics[width=#1in]{./fig/#2}
\end{center}
\caption{#3}
\label{#4}
\end{figure}
}

\def\twofigs#1#2#3#4#5#6#7{

\begin{figure}[t]
	\centering
	\subfigure[#7]{
	\centering
	\includegraphics[width=.45\linewidth]{./fig/#1}
	}
	\subfigure[#6]{
	\centering
	\includegraphics[width=.45\linewidth]{./fig/#2}
	}

	\caption{#4}
	\label{#3}

\end{figure}

}

\AtBeginDocument{%
  \providecommand\BibTeX{{%
    \normalfont B\kern-0.5em{\scshape i\kern-0.25em b}\kern-0.8em\TeX}}}

\usepackage{filecontents}

\begin{document}

\date{}

\title{Extending the Control Plane of Container Orchestrators for I/O Virtualization}

\author{
 Garegin Grigoryan\\ grigoryan@alfred.edu  \\Alfred University 
 \and
 Minseok Kwon \\ jmk@cs.rit.edu  \\Rochester Institute of Technology 
 \and
 M. Mustafa Rafique\\ mrafique@cs.rit.edu  \\Rochester Institute of Technology
}

\maketitle

\begin{abstract}
Single Root Input/Output Virtualization (SR-IOV) is a standard technology for forking a single PCI express device and providing it to applications while ensuring performance isolation. It enables container orchestrators to share a limited number of physical network interfaces without incurring significant virtualization overhead. The allocation of virtualized network devices to containers, however, needs to be more configurable based on the bandwidth needs of running applications. Moreover, container orchestrators' network control over the virtualized interfaces is limited by the abilities 
of SR-IOV. We explore the design considerations for a system with controlled SR-IOV virtualization and present ConRDMA, a novel
architecture that enables fine control of RDMA virtualization for containers. Our evaluation shows that ConRDMA enables containers to use RDMA allocated bandwidth more efficiently and to select best-suited nodes to meet their varying communication requirements.
\end{abstract}

\section{Introduction}

A hardware-based I/O virtualization technology called Single-Root Input/Output Virtualization (SR-IOV)~\cite{sriov} enables zero-copy PCI express device virtualization for applications, while providing isolation between virtual devices. SR-IOV is used for achieving high network throughput when forking a physical NIC into virtual NICs. 
Containers~\cite{docker,coreos,baier2018getting} increasingly become an essential part of multi-tenant clusters like clouds due to their capability to offer isolation, portability, and low complexity~\cite{baier2018getting}. Container orchestrators are used to provide large-scale container deployment and management, such as system and  network configuration. Container orchestrators use SR-IOV for sharing a limited number of network interfaces among multiple containers without degrading network throughput and increasing latency. 

Using SR-IOV for container networking, however, limits the manageability of underlying networking traffic. In this work, we look at the current architecture of RDMA SR-IOV virtualization by Kubernetes \cite{k8s, melhowto} for container management. The Kubernetes plugin for virtualized RDMA~\cite{k8plugin} developed by Mellanox constructs an overlay network among pods (i.e., groups) of containers. 

Unfortunately, the current solution for virtual RDMA lacks in: 1) bandwidth rate-limiting, 2) heterogeneous nodes, 3) multiple interfaces, and 4) per-container Virtual Function (VF) allocation. The ability to customize virtualized interfaces is limited; the configuration map of Kubernetes does not consider different capabilities of VFs at each node; the Container Network Interface (CNI) plugin does not allow attaching more than a single RDMA interface to a pod; and currently, containers, not pods, request interfaces. However, this design does not provide containers' network isolation within a pod of Kubernetes since the interfaces used by containers within a pod are shared. 

As a case study, we extend the control plane of Kubernetes to support containers using virtualized RDMA with fine control. At the heart of our extension called ConRDMA are the following components:
\begin{enumerate}
    \item {\it RDMA Hardware Daemon Set} that runs on each node for initializing RDMA resources and acts as a conduit between Kubernetes and custom resources such as VFs for RDMA;
    \item {\it Scheduler Extender} that finds the most appropriate node for deploying a pod with RDMA requirements;
    \item a modified {\it Container Network Interface (CNI) plugin} that virtualizes an RDMA NIC for multiple containers.
\end{enumerate} 

Kubernetes deploys a new pod to the best node based on the RDMA resource information obtained from the daemon set. The CNI finalizes the process by setting the network interfaces, IP addresses, and moving the VFs from the node's namespace to the pod's. The daemon set addresses the problem of communication between the CNI and the Kubernetes Device Plugin. This lack of communication may cause discrepancies of the physical amount of VFs that Kubernetes believes there are against those are truly allocated in the CNI. 

ConRDMA complies with the Kubernetes standards allowing new additions and extensions to be integrated into the current Kubernetes easily with no conflicts. It also allows for bandwidth limiting for RDMA interfaces, and specifies interfaces on a per pod basis, not container. We used the standard benchmarking tools to evaluate the correctness of bandwidth reservation and the influence of bandwidth control on the latency of RDMA SEND operation. Unlike the existing plugin \textit{k8s-rdma-sriov}, ConRDMA controls throughput based on the needs of a pod application. Moreover, ConRDMA allows pods in a Kubernetes cluster to have more than one virtualized RDMA NIC to be allocated with. 

To summarize, while preserving the advantages of SR-IOV including high throughput and compatibility with the standard Kubernetes and container design, ConRDMA allows fine-grained RDMA interface management and rate-limiting within a heterogeneous cluster.

\section{Background}
\textbf{Container networking.} Datacenters employ virtualized infrastructure such as containers for reduced cost and easy management including server migration, recovery and testing. Containers virtualize the OS that is shared by all the containers, and this sharing makes system management like storage and networking lightweight, efficient, and scalable. In networking, however, containers impose additional overhead as data pass through two network stacks, one at the host and the other at the container. This inevitably results in degraded throughput and latency~\cite{afgraft}. Also, the containerized applications running on a server are decoupled from hardware as well as the OS, making it hard for the NIC to have accurate network flow information. This is because end-to-end flows are broken due to containers, and thus the applications may not be able to benefit from RoCE~\cite{rocev2}, RDMA~\cite{rdma}, let alone Data Center Quantized Congestion Notification (DCQCN)~\cite{dcqcn} that requires transport header analysis. To tackle this challenge, three approaches have been proposed: 1) interface virtualization, 2) optimized network stacks, and 3) socket grafting. 

\textbf{Container orchestration.}
The management of containers becomes challenging at scale, and container orchestration has been introduced to provide efficient management and deployment. It provides automated services like finding the most appropriate host to deploy a container, allocating resources between containers, migrating containers when the original host fails, balancing container loads, and routing between containers transparently. In Kubernetes~\cite{k8s} and Docker Swarm~\cite{swarm}, which are popular container orchestrators, a master node interacts with worker nodes to decide on which node a container should be placed. The worker nodes monitor the resource status, and report information back to the master node. For fault-tolerance, migration, and routing, again the master node communicates with the worker nodes, gathers information, and makes decisions.

\textbf{SR-IOV.} SR-IOV~\cite{sriov_intro} is an I/O virtualization technique that allows PCIe, e.g., network adapters, to be shared in a virtual environment for high performance and manageability.  SR-IOV achieves high performance by bypassing software virtualization stack with little extra overhead introduced in  providing isolation. To achieve this, SR-IOV enables I/O memory management unit (IOMMU) to handle memory translation between the physical and virtualized NICs. While SR-IOV allows each device to provide no more than 256 VFs (Virtual Functions), SR-IOV virtualized devices experience little loss in network bandwidth compared to physical NICs.

\section{SR-IOV with Containers}
Currently, Kubernetes with SR-IOV virtualization allows virtual functions (VFs) to be requested only on a {\it per-container} basis, and this is its primary limitation with significant ramifications. This feature has an unintended consequence that all of the containers within a pod share the network namespace, and thus have access to the same network interfaces. This means that a container not requesting for a virtualized interface can also use the interface simply because the container belongs to the same pod as the one that indeed has requested for the interface. This drawback is partially alleviated as the CNI plugin assigns only one VF to each new pod--this ends up like VF allocation per pod. 

Unfortunately, this technology leads to an even more serious problem--Kubernetes believes that multiple virtual functions are in use by a pod (one or more for every container), while in reality only one is in use. This makes the VF device resources appear (to Kubernetes) to deplete faster than they actually do (because multiple VFs are falsely believed to be in use). This can result in pods whose network requirements could not be satisfied to be deployed due to SR-IOV VF device resources within the worker nodes falsely appearing to be depleted. 

It is also not possible to counter this effect by requesting an RDMA VF in only one container per pod although this would ensure that the number of requested VFs and the number of allocated VFs stayed in sync. The RDMA device plugin provides a privileged file system mount (/dev/infiniband/) to containers that request an RDMA VF device on a per-container basis since the containers within the pod do not share a file system namespace by default. This also implies the containers that do not request a virtual function to be left without the ability to send data on the RDMA interface for their pod.  

Use of virtual functions as a device resource presents another roadblock in an attempt to implement a fined-grained control mechanism in container networking over SR-IOV virtualized interfaces. If pods (or containers) are allowed to reserve bandwidth, it is possible for a single virtual function to reserve all the bandwidth on its corresponding physical interface. This effectively leaves the remaining VFs on that interface unusable though they are device resources. One way to circumvent this is to present available bandwidth as the device resource, rather than VF's, but this only moves the problem to the possible depletion of VFs while there is still bandwidth available.

Further complications arise in the face of multiple physical RDMA interfaces per node. In such a setup, it is tremendously helpful to know available bandwidth on each individual physical interface because a single VF assigned to a pod cannot use bandwidth from multiple physical interfaces, especially when the bandwidth available on its interface is not sufficient.

\section{Design Considerations}

To address the aforementioned problems, we need to control the generation and allocation of SR-IOV-virtualized interfaces and manage their bandwidth limitations within a container orchestrator. To support this, we need to extend the container orchestrator to be able to 1) monitor the status of the physical and virtualized interfaces at every node of the cluster; 2) install pods into selected nodes based on the status; 3) allocate virtualized network interfaces for the installed pods and set them up with respect to pods' specifications.

Below we present components that would be necessary to design such a system:
\begin{enumerate}
    \item A hardware daemon to run on each of the worker machines of the Kubernetes cluster. A hardware daemon set is an extended device plugin, and thus provides resource information about CPU and memory as well.  Most importantly, it tracks the number of active Physical and Virtual Functions and sets up Virtual Functions according to the specification set by an operator of Kubernetes. 
    \item A scheduler extender for Kubernetes, to collect information from the hardware daemons on available hardware resources including networking interfaces. The scheduler extender uses that information to identify nodes with sufficient free bandwidth and virtual functions to fulfill the requirements of the pod.
    \item A Container Network Interface (CNI) for moving virtual interfaces into the namespaces corresponding to pods that were allocated with those interfaces. In addition, CNI plugin is responsible for the assignment of IP addresses to the allocated virtual interfaces.
\end{enumerate}

\section{Case Study: Containers Using RDMA}

\twocolfig{2.7}{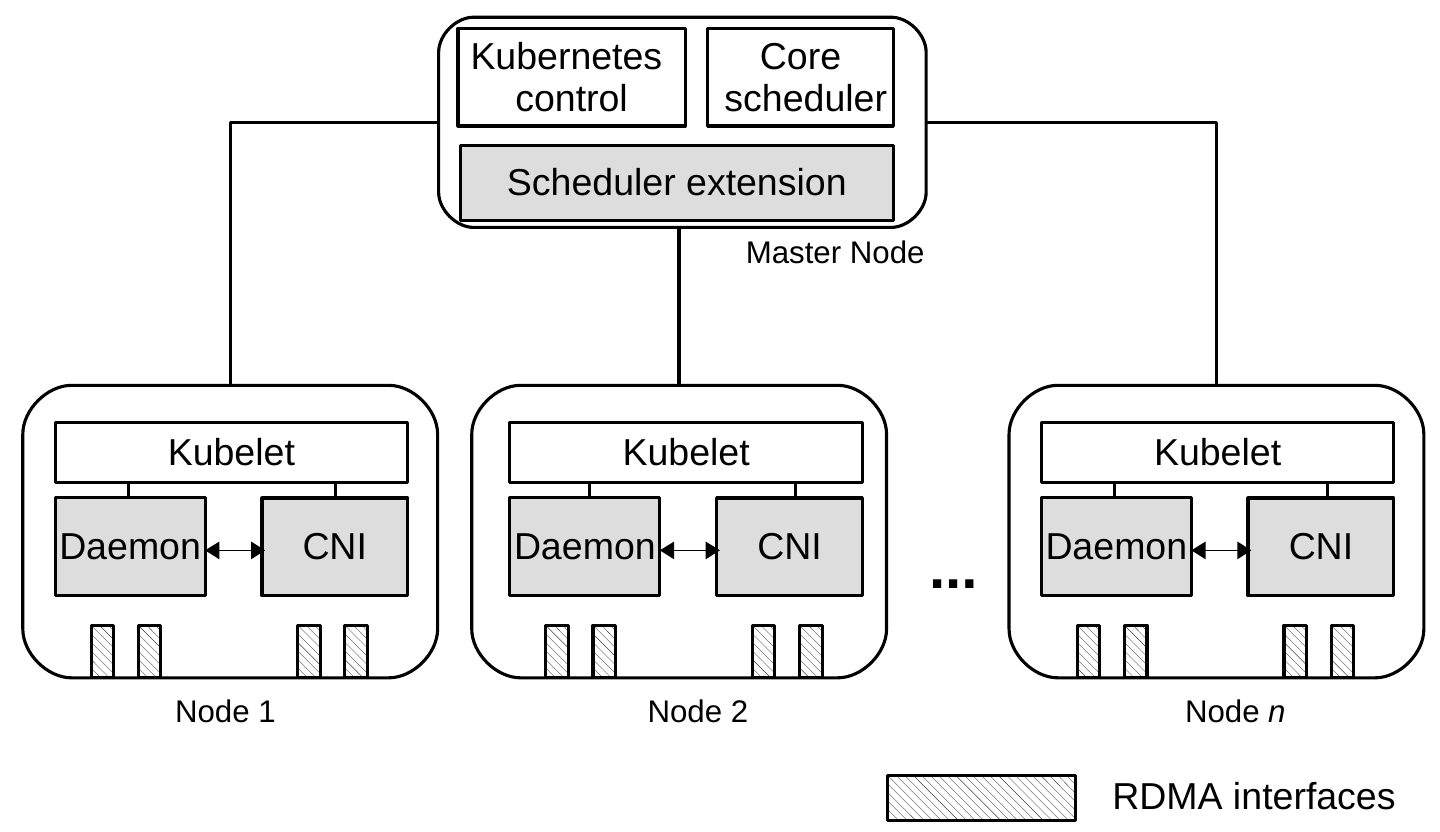}{Design of ConRDMA.}{fig:ext}

We now show how the key components in the design considerations can be implemented in the control plane of container orchestrators through a specific instance. As a case study, we extend Kubernetes Remote Direct Memory Access (RDMA)~\cite{rdma} SR-IOV device plugin to be able to impose limits on pod placement for RDMA resources. RDMA has been developed to bypass the CPU and for network interface cards (NICs) to directly transfer data to and from memory. Our extension in Kubernetes,  called {\it ConRDMA}, is able to support more sophisticated and efficient SR-IOV virtualized interface allocation for container networking via fine-grained virtualized RDMA control. Primarily, ConRDMA enables communication between the Kubernetes scheduler and the monitoring drivers for RDMA hardware resources to help promote the placement of containers (or pods) to well-suited nodes. The dark components on Figure~\ref{fig:ext} show our changes and additions to the existing Kubernetes design for RDMA.   

One critical benefit of ConRDMA is to allocate RDMA VFs on a per-pod basis instead of per-container. This change reflects the underlying setup more accurately as the containers in a pod share the given RDMA bandwidth. Specifically, use of the scheduler extender and RDMA daemon set pairing ensures that the portion of the pod definition that requests RDMA resources to be parsed only by the scheduler extender and the CNI plugin. This makes it straightforward to extend or modify the format of this specification, and allows it to be placed in the annotations section of the pod definition. Additionally, the format and complexity of the RDMA requirements specification is no longer limited by the device resource API, since no core Kubernetes components need to parse the RDMA requirements specification for a pod. Also, the extended architecture has backward compatibility as it can process pods either with annotations that follow the original format or with no annotation about RDMA.

\subsection{Overview of ConRDMA}


\twocolfig{2}{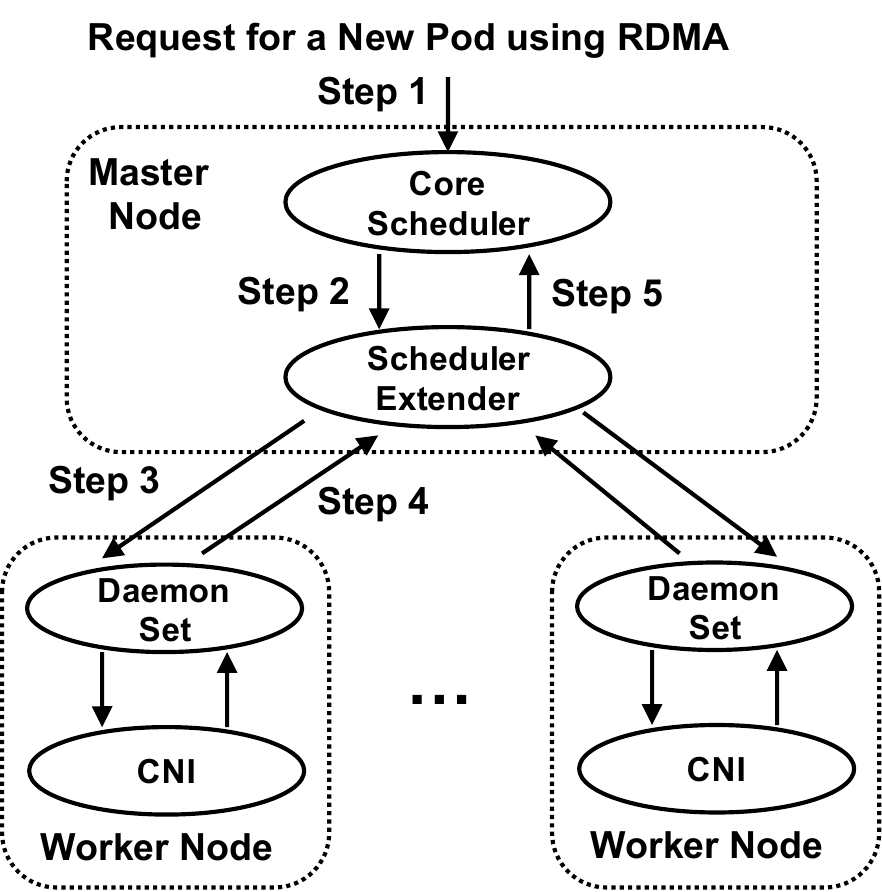}{System architecture of ConRDMA.}{fig:archi}

We extend Kubernetes to allow containers to use RDMA  efficiently and in a more flexible way to meet diverse needs of applications. This extension consists of three steps, namely node selection, CNI information collection, and the creation of VFs, as shown in Figure~\ref{fig:archi}. We add the scheduler extender at the master node, the hardware daemon set at each worker node as new components, and extended the CNI again at worker nodes. We will describe the goal of each step (as shown in the figure), and how to achieve the goal in detail.

First, Kubernetes chooses which node the pod should be deployed on based on available resources including RDMA. Second, Kubernetes receives a request from an application (or users) to schedule a pod with needed number of RDMA interfaces (step 1). Upon receipt of the request, the Kubernetes core scheduler discovers all nodes that meet the core requirements in terms of CPU and memory (step 2). The next step is to identify a subset of the nodes that can provide the required RDMA interfaces by the scheduler extender. The scheduler extender receives the list of nodes from the core scheduler, and then contacts the daemon set running on each node to gather their PF and VF information (step 3). The daemon set returns the metadata information on PFs and VFs back to the scheduler extender (step 4), which in turn computes the information (see Section~\ref{sec:impl} for details) and returns the filtered list of nodes that can host the pod back to the core scheduler (step 5). Finally, the core scheduler determines a valid node from the returned list of nodes and begins placing the pod onto that designated node.

Once the node on which the pod will be deployed has been selected, the CNI gathers information on RDMA interfaces to determine which VFs to be used for the pod. For that, the CNI contacts the daemon set that scans all the available RDMA interfaces. The RDMA information is returned to the CNI that determines which VFs on each interface to be used within the pod. Figure~\ref{fig:seq} is a sequence diagram that shows the function call flow for choosing a node for container deployment.

\twocolfig{3.3}{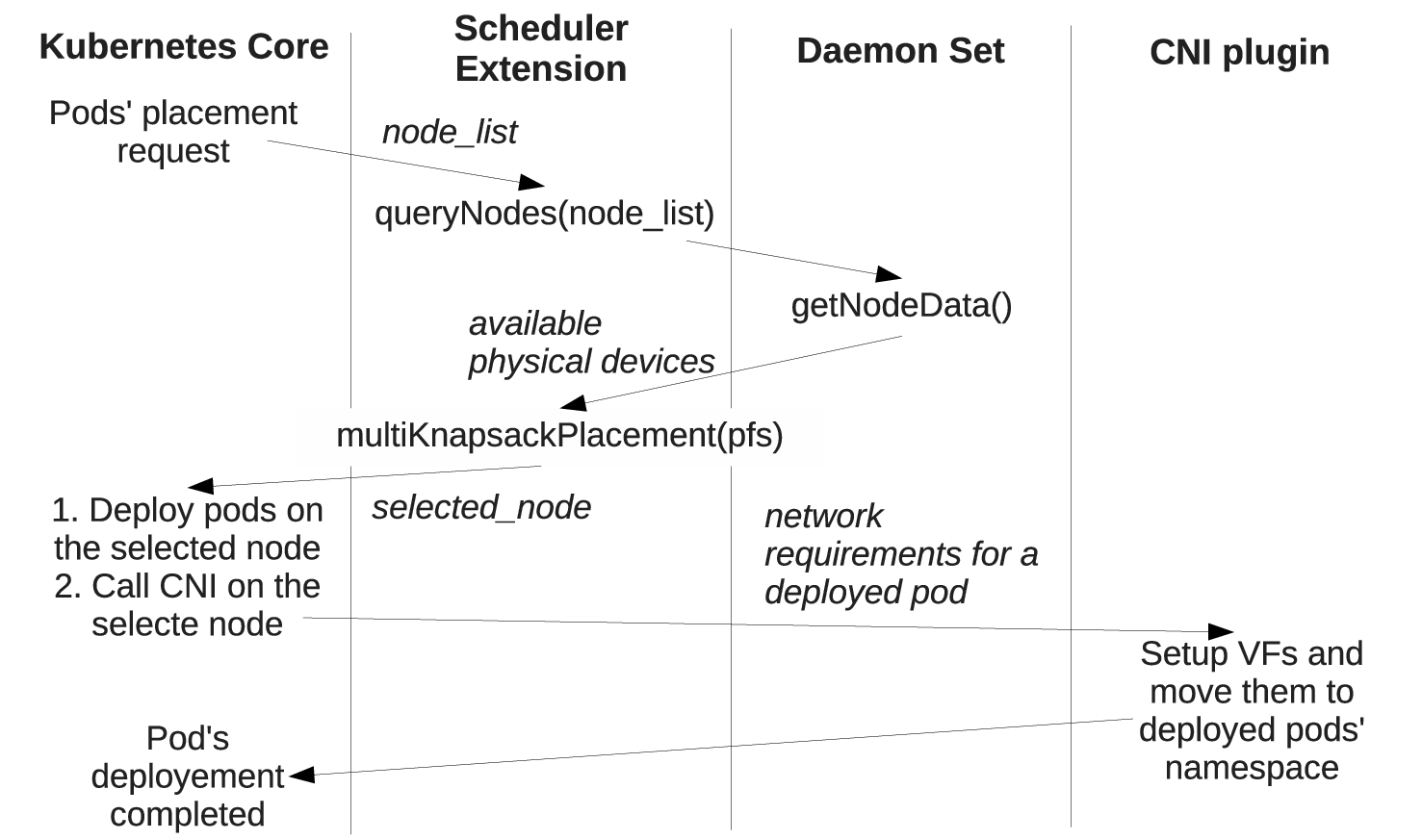}{The function call flow for node selection for container deployment.}{fig:seq}

The goal of the final step is to create the VFs identified by the CNI and allocate them to the pod. The CNI now knows the names of the VFs passed by the daemon set in the previous step and the bandwidth requirements specified in the request in the beginning. It changes the network namespace of the VFs from the node to the pod, binding the VFs to the pod. The CNI then sets the bandwidth limits for each of the VFs within the pods network namespace. Once this completes, the metadata of the network configuration is returned to the Kubelet process that finishes setting up the pod.

Additionally, the CNI handles the case when either a VF fails to set up or when it needs to close a VF when the pod is shut down. In that case, the CNI returns the state of the system back to where it was before the pod initialization, so that other pods can utilize the VFs allocated to the pod. Specifically, the CNI first moves any VFs that are placed on the pod back into the node namespace, and reverts any changes to the VFs such as the bandwidth limitation.

Next, we elaborate on details of ConRDMA's components, specifically, RDMA hardware daemon set, the Scheduler extender and Container Networking Interface (CNI).

\subsection{Implementation}
\label{sec:impl}

The \textbf{RDMA hardware daemon set} (in short, \textit{daemon set}) provides information on hardware setup, initialization, and RDMA resources in use and available for use. A daemon set runs at each node as a pod that consists of two containers: \textit{init} and \textit{server} containers. The init container, which is launched before any other containers in the pod, initializes all the VFs that exists on the node. It begins with scanning all the interfaces on the node, and then determines whether the interface is RDMA- and SR-IOV- enabled. After that, the init server configures all the VFs for those interfaces. This hardware setup and initialization functionality is already provided by the Mellanox RDMA device plugin~\cite{k8plugin}. The server container extends the device plugin with a REST endpoint that communicates information about PFs that contain metadata on the capacity and available RDMA resources. 

The \textbf{scheduler extender} allows decisions on node eligibility to be made based on RDMA information or constraints outside of those that Kubernetes implicitly understands. The scheduler extender is a process registered with the core Kubernetes scheduler as an extension all running on the master node. During the scheduling of a pod, the core scheduler calls out to this extension via HTTP requests, and receives its input about a candidate node(s) in the cluster the pod can be deployed on. The scheduler extender queries each instance of the daemon set to retrieve information on the availability of RDMA VFs on the node, the number of VFs in use, and how much bandwidth has been reserved on each physical interface. 

After collecting this information from each available node, the scheduler extender filters the available nodes to only include those with enough free RDMA bandwidth and VFs to fulfill the requirements of the pod. Specifically, the scheduler extender decides whether the node has the necessary requirements for a pod based on the number of VFs required and the minimum bandwidth requirements of each of those VFs. If a pod has multiple RDMA interfaces, each with a minimum bandwidth reservation, an instance of the {\em multi-knapsack} problem is solved by the scheduler to identify if the node can host the pod. For example, a pod that needs two VFs with a bandwidth of 100~Gb/s each is placed on a node with either a single interface that has at least 200~Gb/s of unused bandwidth or two interfaces that each has at least 100~Gb/s of unused bandwidth. Of course, the scheduler extender places a pod with no RDMA requirement following the original scheduling behavior. When no node satisfies a given RDMA requirement, Kubernetes fails to place the pod and returns an error.

We use \textbf{Container Networking Interface (CNI)} to configure the namespace of a pod for VFs \cite{cni:cni}. CNI is a specification that determines how network interfaces are provided to and removed from containers when a pod is created or torn down. In Kubernetes, the kubelet process invokes CNI to initialize the network interfaces, assign IP addresses, and allocate the selected VFs to the pod.  Which CNI plugin is used can differ from one node to another within a Kubernetes cluster, although a single plugin type is typically used across all nodes--we use the Mellanox CNI~\cite{mel:cni}. The CNI plugin on a node is only executed once during the deployment of a pod regardless of the number of containers that pod contains. This is because the containers in the pod share a network namespace, and thus share the network interfaces inside of it.

We extend the CNI as follows. During the startup, the CNI reads the pod configuration, and determines the number of VFs that have been requested. The CNI then decides which interfaces should be used based on the VF information, and moves all the requested VFs to the network namespace of the pods from the nodes network namespace. Consequently, the interface name of each requested VF is changed to \textit{eth[num]}, where \textit{num} is an integer starting at 0 and incremented for every VF added. Now that the VFs are moved successfully and have proper IP addresses, the CNI sets up the bandwidth requirements for every requested VF via the \textit{/sbin/ip} command. To shut down a pod, all the VFs of the pod are moved back to the original namespace of the node, their interface names are rolled back, and any bandwidth configurations that were added are removed.


\section{Evaluation}
We first focus on bandwidth allocation for containers with or without ConRDMA, and then evaluate how the node selection algorithm performs. We conducted experiments on three nodes with a master node (an Intel Core i5-2400 machine with quad cores), and two worker nodes, each with Intel Xeon CPU E5-2640 with 20 cores. The worker nodes were provided with two Mellanox ConnectX-5 RDMA NICs, and were connected to each other.

\subsection{Bandwidth Allocation}
We used the benchmarking tools \textit{ib\_send\_bw} and \textit{ib\_send\_lat} from the \textit{perfest} package of Mellanox~\cite{perftest} for evaluating RDMA bandwidth control by ConRDMA and its impact on the two-sided SEND operation of RDMA. We deployed three pairs of containers with containers from each pair installed on the two separate nodes. Each node has a single RDMA interface of a maximum bandwidth of 100 Gb/s. The containers were allocated with the minimum RDMA bandwidth as follows: (a) video streaming containers for 60 GB/s, (b) AI-node containers for 30 Gb/s, and (c) file storage containers for 10 Gb/s. We measured the maximum bandwidth between the containers of each pair to verify the rate limiting capabilities by ConRDMA, and compared the performance with the original Kubernetes-RDMA design without fine-grained control. We first ran data transfer between the first pair of containers, i.e., videostreaming containers, and then  simulated AI data transfer between the second pair of containers. After that, we simulated the data transfer between file storage containers. Finally, we gradually stopped the transfer between the videostreaming, AI and file storage containers.

\twofigs{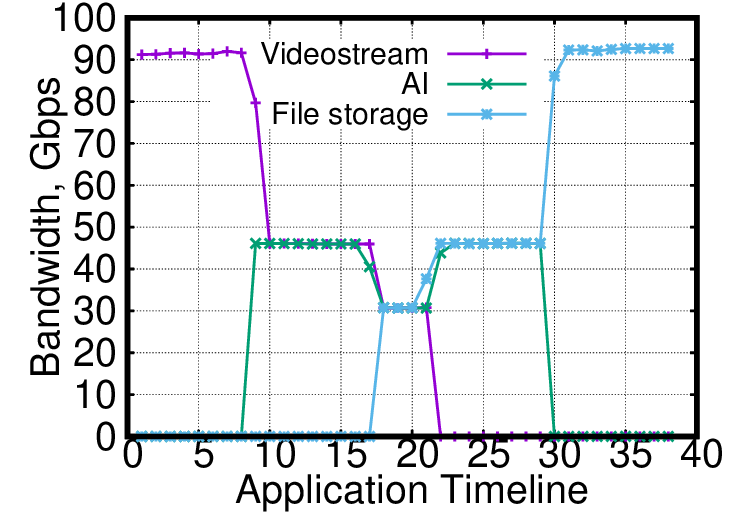}{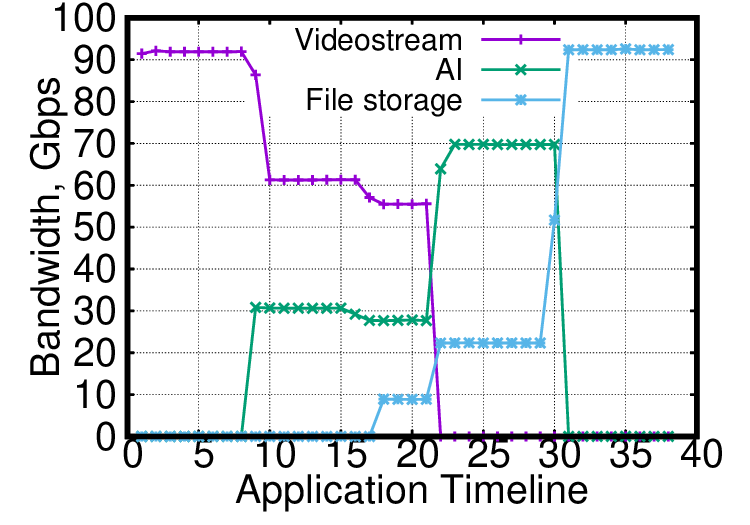}{fig:bandwidthcontrol}{RDMA bandwidth control in Kubernetes.}{1.6}{Bandwidth control enabled.\label{fig:bl}}{Bandwidth control disabled.\label{fig:nobl}}

Figure~\ref{fig:nobl} shows that the containers share RDMA bandwidth equally without bandwidth control. In contrast, ConRDMA (see Figure~\ref{fig:bl}) allows regulating the bandwidth of containers based on their needs. For example, the videostreaming containers are allocated the largest share of RDMA bandwidth, as it was specified during their deployment--at the 17th iteration when running simultaneously,. The file storage containers receive the smallest share, as their minimum bandwidth requirement is lower than that of the AI containers. As soon as the other flows stop, the file storage containers regain the full share of throughput (after the 30th iteration).

\begin{figure}
     \begin{minipage}[t]{.48\linewidth}
                \includegraphics [width=\linewidth]{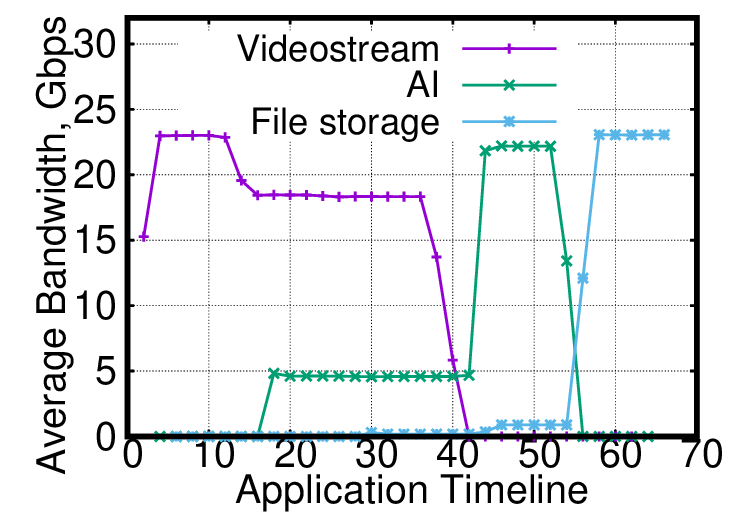}

                \caption{RDMA bandwidth control for multiple pods.}
            \label{fig:multiple}
            \end{minipage}
            \begin{minipage}[t]{.48\linewidth}
                \includegraphics [width=\linewidth]{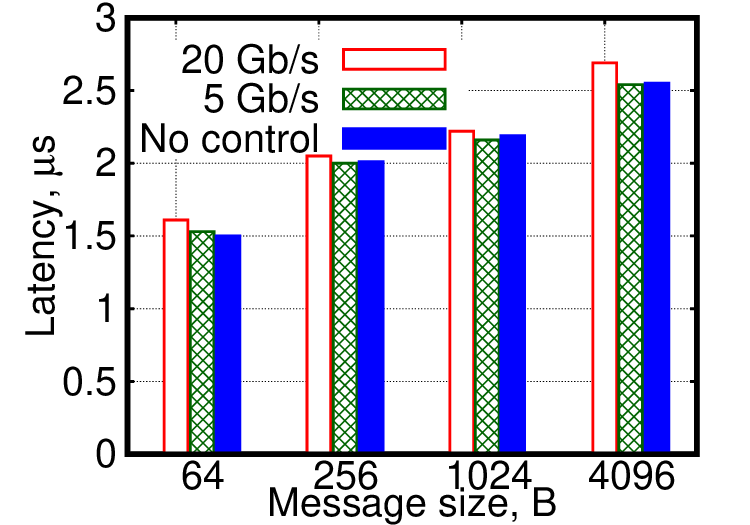}

                \caption{Latency and RDMA bandwidth control.}
            \label{fig:latency}
            \end{minipage}

\end{figure}

We also evaluate the performance when even more pods are deployed at the same node. We installed four pods of each application type (videostreaming, AI, and file storage) at each node. While the  videostreaming and AI pods were set to the minimum bandwidth of 20 Gb/s and 5 Gb/s, respectively, the file storage pods were not assigned any bandwidth requirement. Figure~\ref{fig:multiple} shows that ConRDMA behaves as expected, i.e., correctly allocates bandwidth with respect to the configured requirements. 

We verify that ConRDMA does not significantly increase network latency by running a \textit{ib\_send\_lat} benchmark test. We measured  round-trip time between two hosts with and without the minimum bandwidth allocation. Our results show that the minimum bandwidth allocation has little effect on latency (see Figure~\ref{fig:latency}).

We also observe: either of the flows did not use the maximum amount of bandwidth allowed, and even when there is enough bandwidth, the flows share it proportionally, not equally, according to their minimum bandwidth needs (see iterations 21-30 in Figure~\ref{fig:bl}). ConRDMA does not directly control bandwidth allocation, but rather uses the \textit{/sbin/ip} binary and Mellanox drivers to set limitations on each virtual function. 

\subsection{Node Selection}
ConRDMA places containers based not only on RDMA resources, but available RDMA bandwidth. Consider a cluster with two nodes, each with two RDMA interfaces with maximum bandwidth of 100 Gb/s. In such an environment, we deployed a videostreaming container \textit{A} with the requirement of two interfaces with minimum 80 Gb/s bandwidth each. Next, we deployed an AI container \textit{B} with  two interfaces with minimum 50 Gb/s each. Finally, we deployed a container \textit{C} for file sharing with two interfaces, minimum 30 Gb/s bandwidth each. Without the rate limiting capability, containers \textit{A} and \textit{C} are placed on the same node. With ConRDMA, however, container \textit{A} with a minimum bandwidth requirement of 80~Gb/s is always deployed on a separate node, since containers \textit{B} or \textit{C} have the minimum bandwidth requirement more than 20~Gb/s.

It should be noted that ConRDMA rejects pod installation if a required minimum bandwidth is not guaranteed for a candidate pod. We leave an exploration of more intelligent pod placement algorithms that minimizes the rejection rate for future work.


\section{Related Work}
Modern distributed systems, such as data centers and clouds, require fast transfer of large blocks of data across the data center network. The standard packet processing at end hosts involves significant CPU overhead, as each packet needs to be copied from the Network Interface Cards’ (NICs) buffers and passed through the OS kernel~\cite{guo, kim2018hyperloop}. 
Remote Direct Memory Access (RDMA) via hardware NICs, such as Mellanox cards~\cite{mellanox}, enable zero-copy packet processing and minimize CPU involvement. Initially RDMA was designed for InfiniBand communication standard. However, most of the data centers' networks' link layer relies on Ethernet, which spurred emergence of protocols such as RoCE~\cite{beck2011performance} and RoCEv2~\cite{rocev2}. Both protocols enable RDMA through Ethernet network. While RoCE is designed for Ethernet networks within the same broadcast domain, RoCEv2 operates at the Layer 3 (network layer) and encapsulates RDMA in routable IPv4/IPv6 packets. Lu et al. in~\cite{lu2018multi} give concise background on the architecture of RDMA.

To provide high network throughput and both flexibility and reliability in the system, data center operators are highly interested in integrating RDMA networks with containerized system~\cite{liss2015containing}. The main challenge for such an integration is that RDMA is a hardware solution and needs to be virtualized on order to be used by multiple containers simultaneously. There are several ways for such a virtualization. 
SR-IOV~\cite{sriov_intro} is a standard technology for forking a single PCI express device and providing it to applications, while ensuring the isolation. Mellanox provides an SR-IOV handler~\cite{mellanox} for Kubernetes, an open-source system for automating deployment, scaling, and management of containerized applications~\cite{k8s}. vRDMA~\cite{vrdma} virtualizes RDMA interface for VMware guest machines, and enables resource management for the VMware hypervisor. FreeFlow~\cite{freeflow} uses virtual switches to bridge between RDMA and containers, and to coordinate datapath routing and container placement. 

\section{Conclusion}

The current SR-IOV solution for container orchestrators lack in bandwidth rate limiting capabilities, and we argue that the control plane of container orchestrators needs to be extended to enable this. Specifically, we implement the hardware daemon, Container Networking Interface plugins, and the scheduler extender. As a case study, we design ConRDMA, a system that allows RDMA virtualization in a Kubernetes cluster with improved manageability. We tested ConRDMA in a real system with Kubernetes, Docker containers, and Mellanox RDMA. Our results show that ConRDMA promotes accurate bandwidth allocations close to user requests.

\section{Future work}

The key question that we seek in this work is whether we can design and implement an I/O virtualization technique that can be easily used by heterogeneous containers and their orchestrator, and we appreciate feedback from the workshop audience. We use RDMA with SR-IOV as a specific I/O virtualization technique. It is important in providing such a technique to improve its controllability while not compromising isolation, portability, and the RDMA performance. The primary challenge in doing so is how to decouple virtual RDMA resources from the container orchestrator,
so that heterogeneous servers and the orchestrator can use them without depending on constraints by other components. To achieve our goal and address the challenge, we consider the following three research questions as future work:

\begin{enumerate}

\item {\it Can we develop an I/O virtualization technique for flexible use by containers?} 
This will require the following three components 1) a backend module (residing at the physical host), 2) a front end module (at the container), and 3) a device driver (possibly a driver plugin). The backend module monitors I/O resources and reports this information to the front end module that places containers to eligible hosts. The device driver allocates resources (RDMA in our case) at the NIC for the front end module to launch containers in the relevant namespace.

\item  {\it Can we extend container orchestrators into Software-Defined virtualized I/O management?} 
We plan to develop Software-Defined I/O management by expanding the container orchestrator for containers to use virtualized I/O efficiently in a user-defined way. Currently, it is not clear how to use virtual I/O (e.g., RDMA) in the SDN setting, especially for containerized applications. One approach would be to have a control plane, the user plane, and the communication protocol between them.

\item {\it How does the control plane of container orchestrators enable efficient container placement and migration?}
We will need to implement algorithms for container placement, migration, and consolidation, probably based on some sort of bin packing strategy or the knapsack problem, to achieve this with variable and dynamically changing I/O resources and different workloads. Consolidation aims at maximizing the number of nodes with no virtual machines running on them for making more nodes available for future virtual machines or energy saving (those virtual machines can go to sleep mode).

\end{enumerate}

\section*{Acknowledgment}

The work is supported by the Cisco Research Center. The authors are grateful to John Marshall and Warren Carithers for their help. The authors would also like to thank Jesse Saran and Coleman Link for their contribution to the initial design and implementation.

\balance
\bibliographystyle{abbrv}
\bibliography{sample-base}
\end{document}